\documentclass[aps,twocolumn]{revtex4-2}
\usepackage{amssymb,amsfonts,amsmath,bm}
\usepackage{color}
\usepackage{graphicx}
\usepackage{cancel}
\usepackage{subfigure}
\usepackage{amssymb, latexsym, amsmath}
 \usepackage[T2A]{fontenc}

\begin{document}

\title{'Dancing' two-dimensional  patterns in nematics above the flexoelectric instability threshold.}

\author{E.S. Pikina$^{1}$}

\affiliation{$^1$ Landau Institute for Theoretical Physics, RAS, \\
	142432, Chernogolovka, Moscow region, Russia, \\
}

\begin{abstract}

Various types of electrohydrodynamic instabilities in liquid crystals (LCs) are a recurring theme in LC research.
The nonlinear stage of the flexoelectric instability in nematics remains the least studied aspect of this phenomenon.
This work highlights a key advancement in nonlinear physics, providing a compelling example of nonlinear dynamical system behavior. By solving the  system of  nonlinear equations for a nematic in an alternating external field, we derive  from first principles the unique "dancing" behavior of truly  two-dimensional patterns above the flexoelectric instability threshold. Our analysis is restricted to liquid crystals with negative dielectric permittivity and electrical conductivity anisotropies. The resulting solution represents a stable limit cycle near above the Hopf bifurcation, valid within the parameters range where the solutions characteristic exponent is complex. The generalized two-dimensional Ginzburg-Landau-like phenomenological theory of the phenomenon  is developed.

\end{abstract}

\maketitle

For more than fifty years, the physics of electrohydrodynamic instabilities in liquid crystals under applied external electric field has been a subject of research interest \cite{Meyer69,Pikin,Barnik1978,PI91,Pikin1979,RB91,BK96,KP08,PK18,Pikin2018,lavrentovich1,lavrentovich2,lavrentovich3,dierking1,dierking4,dierking5,PMKL26,PK24,PM23,WI63,HZ68,HE69,ML77,RS88,GP89,RS89,RB91,BRS91,PK24,PM23,ZI91,CH92,GP93,BC94,BK96,KH96,KL03,OP05,ZA06,CG09,TK14,KD15,ES16,pof1,pof2,rev1,
Park25,Cous24,Shio20,Lavr21,Gart21,Lee17}.
We emphasize that, since nematic liquid crystals (NLCs) are soft-matter systems, nonlinear phenomena constitute one of the central research directions in the study of nematics.
Accordingly, the formation of spatiotemporal patterns of nematic director distortions under the action of an external electric field has remained an active subject of investigation for several decades \cite{WI63,HZ68,HE69,ML77,RS88,GP89,RS89,PI91,Meyer69,Pikin,Barnik1978,Pikin1979,Pikin2018,RB91,BRS91,PK24,PM23,ZI91,CH92,GP93,BC94,BK96,KH96,KL03,OP05,ZA06,CG09,TK14,KD15,ES16,pof1,Cous24,Shio20,Gart21,Leask25}. Nevertheless, a complete nonlinear description of the phenomenon is still lacking.

We focus our study  on flexoelectric instabilities in nematic liquid crystals, which restricts the analysis to materials with negative anisotropies of both the dielectric permittivity and the electrical conductivity, that is, $(- -)$  NLCs in de Gennes classification \cite{GP93}.
In our previous works \cite{PM23,PMKL26}, we formulated the complete set of the nonlinear dynamic equations for NLCs in an external alternating electric field. Specifically, in  \cite{PM23}, we derived the exact system of nonlinear equations for all soft macroscopic variables associated with the general conservation laws and symmetry breaking, namely, the components of the nematic director $\bm n$, the hydrodynamic velocity components $v_i$, the electric displacement field $\bm D$, pressure $P$ in the regime of sufficiently low electrical conductivity $\sigma$.

In \cite{PK24}, we provided a linear stability analysis of the general system and identified the known bifurcations of the initially uniform director field leading to  static one-dimensional stripe structures or oblique rolls, as well as the first reported Hopf bifurcation leading to stationary time-oscillating two-dimensional structures, which had no theoretical justification for the system considered.
Our recent work \cite{PMKL26} was, to our knowledge, the first nonlinear investigation of the structures formed in nematic liquid crystals under an external alternating electric field slightly above the flexoelectric instability threshold discussed in \cite{PK24}.
We considered the low-Reynolds-number regime and adopted the limiting cases in which temperature (or entropy) can be excluded from the analysis.
In that study, one-dimensional static and dynamic patterns and their stability were investigated.

 The present work reports  two-dimensional solutions of the  nonlinear electrohydrodynamic equations for nematics derived in \cite{PM23,PMKL26}.
 We consider the standard  geometry with nematic in flat cell of thickness $d$ along the $z$-direction powered by an alternating electric field  perpendicular to the $x-y$ plane of the cell.  We assume that without the external electric field unperturbed nematic director $\bm n$ is aligned everywhere along the $x$-axis. We study  external electric field amplitudes within a finite interval  above the flexoelectric instability threshold  for the case where the complex increment (i.e. Lyapunov characteristic exponent or growth rate) of the critical mode has a nonzero imaginary part.
The work addresses which solutions remain stable against finite-amplitude perturbations as well as noise.
It is of interest to determine which solutions possess the widest basin of attraction across the different regions of the phase diagram obtained from the linear analysis.
Addressing these questions requires the investigation of two-dimensional solutions of the nonlinear equations describing a nematic in an external electric field, subjected to various initial conditions with and without noise.

 In \cite{PMKL26}, we provided a detailed description of the reduction from the three-dimensional to the two-dimensional formulation, based on the fact that in the regime $q_c d\gg 1$ the derivatives along the $z$-direction are much smaller than those in the $x$- and $y$-directions. Here,  ${\bf q}_c$  is the finite critical wave vector at which the flexoelectric instability first develops.
This is important since linear analysis shows that the characteristic exponent $\lambda$  contains a contribution proportional to $q_z^2$, alongside contributions proportional to $q_{с x}^2, q_{с y}^2$ \cite{PK24}. The condition $q_c d\gg 1$ enables one to ignore boundaries, i.e. we work in the limit of infinite nematic film thickness.  The linear analysis reveals that in this case no instabilities occur at finite $q_z$, implying the absence of corresponding  degrees of freedom. This justifies the assumption that $z$-dependent unstable modes are absent close to the instability threshold, thereby allowing us to restrict our numerical simulations to two dimensions.

The procedure for eliminating the nonlocal character of the equations for the electric displacement field $\bm D$ and pressure $P$  is also described in \cite{PMKL26}.
This is conveniently achieved by replacing the nonlocal equation for the electric displacement field $\bm D$  with two local dynamical equations for the free charge density $\varrho_e$ and the electric potential $\Phi$, and by introducing fictitious dynamical equations for $\Phi$ and $P$, see below.
Note that the electric displacement field $\bm D$ is expressed as
$D_i = \epsilon_0\, \epsilon_{ik}E_k +  P_{fl, i}$, where $\epsilon_{ik}=\epsilon_\perp \delta^{\perp}_{ik}
+\epsilon_\parallel n_i n_k$ is the permittivity tensor, $\delta_{ik}^\perp\equiv \delta_{ik}-n_in_k\,$, and  the flexoelectric contribution to the polarization is $\bm P_{fl}= \zeta\, \bm n (\nabla \bm n)$.

A comprehensive analysis of the one-dimensional solutions of this nonlinear system for nematics driven by an external electric field of amplitude $E_{0}$  slightly above the flexoelectric instability threshold  $E_{0 c}$ was carried out in \cite{PMKL26}.
In the present work, we  discuss the  two-dimensional numerical solutions of the following system of nonlinear equations:
\begin{eqnarray}
\partial_t n_i =
- \bm v \nabla n_i +n_k \delta_{ij}^\perp \partial_k v_j
+ \, \Xi_i^\perp/\gamma \,
 , \
\label{trun4} \\
\rho \partial_t v_i =  \,
- \nabla \left(K\nabla n_j \partial_i n_j\right)
- \partial_k(\Xi_i^\perp n_k) \,
\nonumber \\
-\, \partial_i P + \partial_k \left(D_k E_i\right)
+ \eta \nabla^2 v_i \, , \
\label{trun3} \\
\partial_t \varrho_e = -\,\nabla \,(\hat \sigma\bm  E) \, ,  \
\label{ro} \\
 \varrho_e = \, \nabla \big(\epsilon_0 \hat\epsilon (\bm E_0 - \nabla \Phi)+ \bm P_{fl} \big) \ ,
\label{pot}\\
 \nabla^2 P + \partial_k\partial_i \left(K\partial_k n_j \partial_i n_j\right)
\nonumber \\
+ \,\partial_k\partial_i(\Xi_i^\perp n_k) - \partial_k\partial_i \left(D_k E_i\right) \,=0\, , \
\label{pres}
\end{eqnarray}
where $\Xi_i^\perp = \delta^\perp_{ij}\Xi_j$, $\gamma$ is the rotational viscosity coefficient, the matrix $\hat \epsilon$ is the permittivity matrix  $\epsilon_{ik}$ of the nematic,
 $\Delta \epsilon=\epsilon_\parallel-\epsilon_\perp$ is the dielectric anisotropy, the matrix $\hat\sigma $ is the electric conductivity matrix   $\sigma_{ik}$ of the nematic, $\Delta \sigma=\sigma_\parallel - \sigma_\perp$ the conductivity anisotropy, and the approximation of a single flexoelectric coefficient $\zeta=e_1$ is used,
\begin{eqnarray}
\hskip-0.0true cm \Xi_i=K \nabla^2 n_i
+ \epsilon_0 (\epsilon_\parallel-\epsilon_\perp) E_i \bm n \bm E
+\zeta E_i(\nabla \bm n)-\zeta\partial_i(\bm n \bm E)   \,
\label{trun6}
\nonumber
\end{eqnarray}
is the negative  variational derivative of the thermodynamic potential over $\bm n$,
 $\bm j=\hat \sigma \bm E$, $n_x=(1-n_y^2-n_z^2)^{1/2}$,  and the total electric field is expressed as
$\bm E = \bm E_0 \cos(\omega t) -\nabla \Phi$,
where $\bm E_0$  is the external electric field directed along the $z$-axis and  $\Phi$  is the potential of the induced distortion electric field.

In the next step, the additional terms  $\delta^\perp_{ik} A_k (\bm A \bm n)$ proportional to $\bm A$, where $\bm A$ is a small vector oriented along the $x$-axis, are introduced into the right-hand side of Eqs. (\ref{trun4}) to keep the predominant orientation of the director $\bm n$ along the $x$-axis and  to suppress  parasitic zero-mode effects in  two-dimensional limit, see \cite{PMKL26}.
Similarly, we add the terms $\, - G_v\,v_i$ to Eqs. (\ref{trun3})  to suppress the zero mode of the velocity components $v_i$.
	These additional terms, proportional to the coefficient $A$, must be small compared with the main contributions proportional  to $\partial^2_x$ and $\partial^2_y$, respectively.

 Note also that in (\ref{trun3}) only  a single isotropic shear viscosity is used instead of the full Leslie viscosity tensor.
The complete viscosity tensor would influence the quantitative values of the instability threshold.
Nevertheless, since the instability is primarily governed by the coupling between the director and the electric field, the set of stationary solutions is expected to remain qualitatively unchanged.

After eliminating the nonlocality of the equations for the electric displacement field $\bm D$ and pressure $P$, see \cite{PMKL26}, one arrives at the following local equations for these fields:
\begin{eqnarray}
{\mathcal{C}} \partial_t \Phi  =  -\,G_1 \Phi +  \varrho_e
-\, \nabla \big(\epsilon_0 \hat\epsilon (\bm E_0 - \nabla \Phi)+ \bm P_{fl} \big) \ ,
\label{potdyn2}\\
{\mathcal{C}_P} \partial_t P= -\,G_P P + \nabla^2 P + \partial_k\partial_i \left(K\partial_k n_j \partial_i n_j\right)
\nonumber \\
+ \,\partial_k\partial_i(\Xi_i^\perp n_k) - \partial_k\partial_i \left(D_k E_i\right) \ , \
\label{trunp1}
\end{eqnarray}
 and
 the following conditions must be satisfied: ${\mathcal{C}} \omega \ll \, \epsilon_0\, \epsilon  \, q_c^2$, $G_1\ll\epsilon_0\, \epsilon  \, q_c^2$, ${\mathcal{C}_P} \omega \ll \,q_c^2$, and $G_P\ll\,q_c^2$.
For the two-dimensional simulations the additional parameters introduced    are chosen as:  $A^2 = 2.5  \cdot 10^{10}\,$ V$^2$ m$^{-2}$, $G_1 = \, 1.54 \cdot 10^{11} \, m^{-2}$. All  additional  parameters satisfy the necessary inequalities discussed above.

In what follows, we briefly discuss the  two-dimensional numerical solutions of the  system of nonlinear equations
(\ref{trun4})--(\ref{ro}),(\ref{potdyn2}),(\ref{trunp1}) over a finite range of field amplitudes  $E_{0}$.
Equations are solved numerically using standard routines realized in Wolfram Mathematica.
 The computational box is chosen to be $\{(- 2\pi/q_{cx}, 2\pi/q_{cx}),(-2\pi/q_{cy}, 2\pi/q_{cy})\}$, where $q_{cx},q_{cy}$ are the components of the critical wave vector,  obtained for the studied set of material parameters within the framework of the linear approximation.

\begin{widetext}

\begin{figure}
{\centering \hskip-0.6true cm
 \includegraphics[width=0.8\columnwidth]{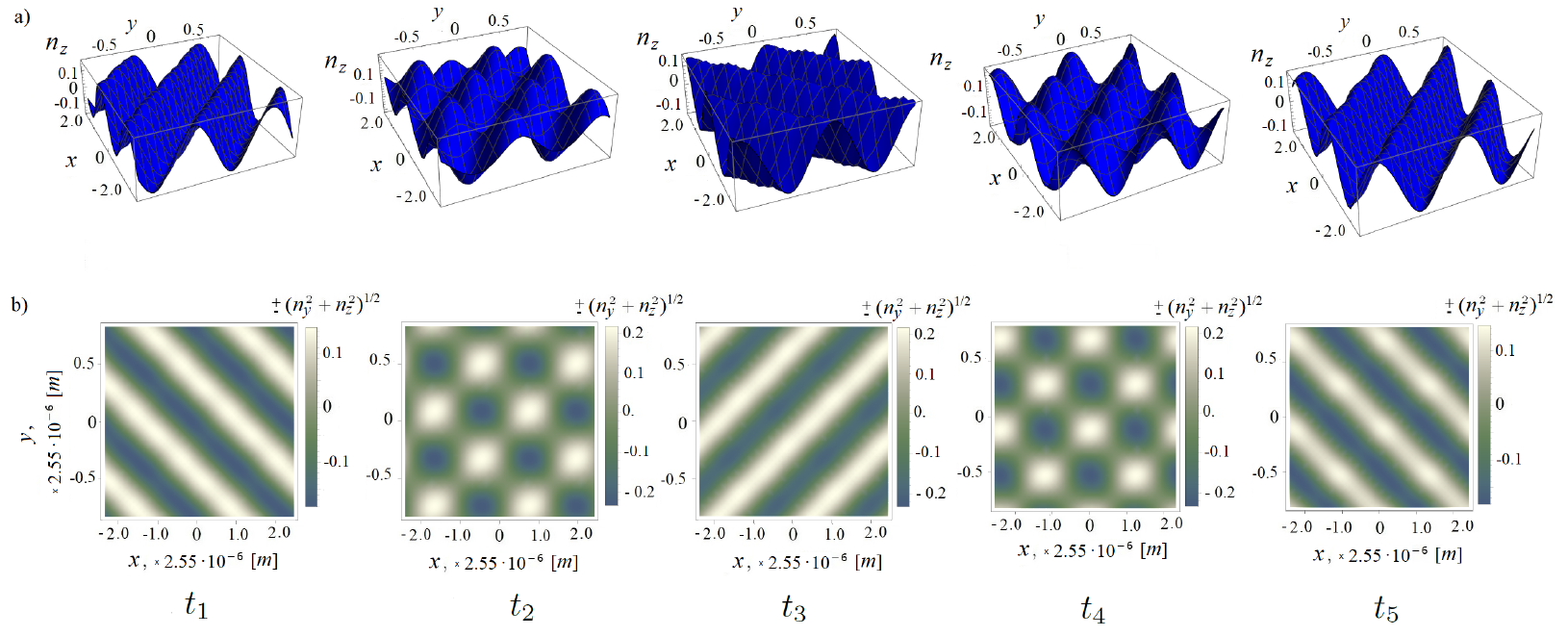}
	 }
\caption{a) Dependence of the distortion field  $n_z$  and  b) density plot of the director distortion field $\vert \delta {\bm n} \vert = (n_y^2 +n_z^2)^{1/2}$ as a functions of the spatial coordinates $(x,y)$  at three five principal time instants during time interval $[0, T_{slow}/2]$  for the solution in the form of the nontrivial two-dimensional  standing wave, which is the stable solution in the range of material parameters where the instability increment of the critical mode has an imaginary part.  Parameters values: $E_{0 c}= 34.8 \cdot 10^{5}\,$ V m$^{-1}$  for the  set of  material parameters:
		$ \omega $ (frequency) $ = 2\pi \cdot\,300\,$ s$^{-1}$;
 		$\sigma_\perp = 20 \cdot 10^{-10}$ $\Omega^{-1} \cdot m^{-1}$, $\Delta \sigma =  - 0.1 \sigma_\perp $,
		$K_1 = K_2 = K_3 = K = 5.5 \cdot 10^{-12}\,$ N,
		$\zeta = 6.75\cdot  10^{-11}$ ${C}\cdot m^{-1}$,
		${\epsilon}_\perp = 18 $,
		$\Delta {\epsilon} = - 4\,\pi \cdot 10^{-1}$,
		$\gamma = \, 0.06\, Pa \cdot s$, $\eta = 0.045\,Pa \cdot s$, $\mathcal{C} = 257.027 \cdot 10^4\, s \cdot m^{-2}$, $\mathcal{C}_p = 102.808 \cdot 10^4\, s \cdot m^{-2}$, $G_p = \, 6.17 \cdot 10^{10} \, m^{-2}$, $G_v =2.467  \cdot 10^{10} Pa\cdot s\cdot m^{-2}$, $q_{c x} = 10.205 \cdot  10^{5}\, m^{-1}$, $q_{c y} = 28.26 \cdot  10^{5}\, m^{-1}$.}
\label{psizdens}
		\end{figure}
\begin{figure}
	{\centering 	\hskip-0.2true cm	\includegraphics[width=0.8\columnwidth]{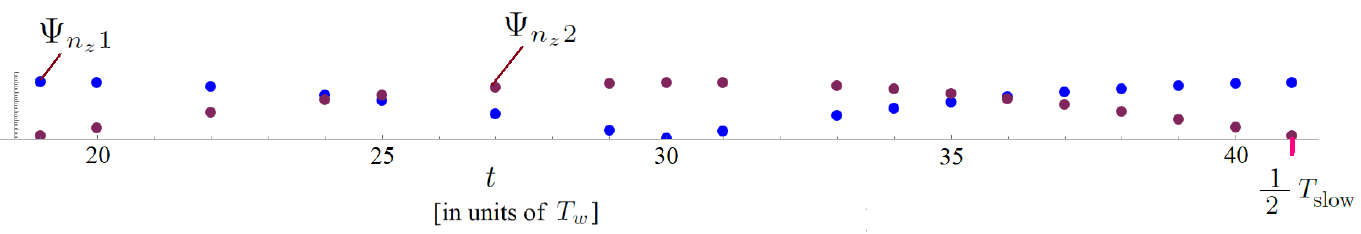} }
\caption{Dependencies of the  absolute values $\Psi_{n_z 1}$ and $\Psi_{n_z 2}$ of the  $n_z$  principal harmonics temporal  prefactors  on  time during  the time interval from the Fig. \ref{psizdens}.}
\label{psiz}
		\end{figure}
\begin{figure}
{\centering \hskip-0.6true cm
 \includegraphics[width=0.9\columnwidth]{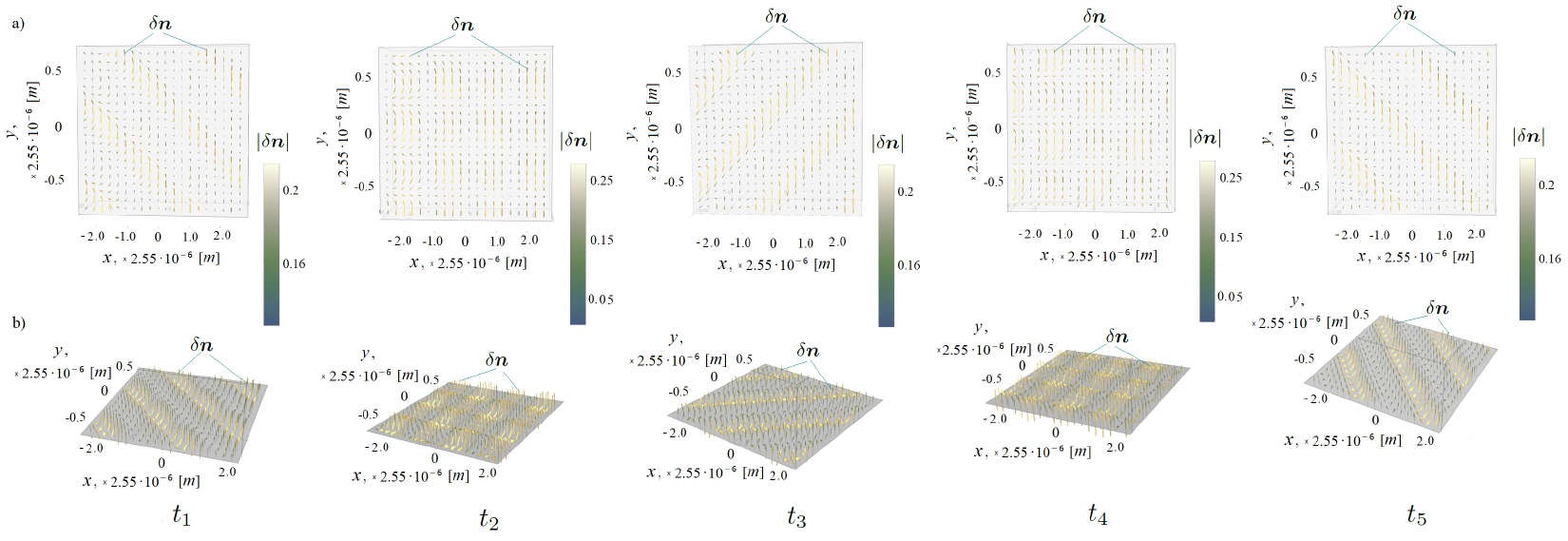}
	 }
\caption{Vector visualization  of the director distortion field  $\delta {\bm n}$: a) top view   and  b) side-view  at three five principal time instants during time interval $[0, T_{slow}/2]$  from the Fig. \ref{psizdens}.}
\label{viz}
		\end{figure}

\end{widetext}

The critical mode has long relaxation times near the instability point.
Thus, the subject of the investigation is the slow dynamic behavior of the NLC on time scales much longer than the period $T_w=2\pi/w$ of the external electric field; here, $w$ is the field frequency.
Thus, the stationary behavior of the solutions occurs on a time scale much larger than the period  $T_w$ of the external field.
The period of the slow dynamics is approximately determined by the value of the imaginary part of the characteristic exponent $\lambda_c$ and is expressed  in units of $T_w$ as $T_{slow} \approx 2 \pi/\mathrm{Im}(\lambda_c)$. In turn, $\lambda_c$ can be obtained within the framework of the linear approximation \cite{PK24}.

 From the numerical solution of the set of the  nonlinear dynamic equations for NLCs (\ref{trun4})--(\ref{ro}),(\ref{potdyn2}),(\ref{trunp1}), we find that when the material parameters and external field amplitude values correspond to the region of the oscillating regime in the linear approximation, then the solution in the form of a nontrivial two-dimensional standing wave is stable.
 Mathematically, this solution reduces to a linear combination of two orthogonal, one-dimensional standing waves
 corresponding to oblique-roll patterns with wave vectors $\bm q_{c 1}=({q_{c x}, q_{c y}})$ and $\bm q_{c 2}=({q_{c x}, - q_{c y}})$, and with a phase shift of the waves $\phi_+= \phi_- = \pm \pi/2$,  see below.
 This regime is illustrated in Figs.  \ref{psizdens},\ref{viz}.
 Fig. \ref{psizdens} shows the $n_z$-distortion field distributions and a density plot of the director distortion field $\delta {\bm n}=(n_y,n_z)$ in the $x,y$ plane at five principal time instants $t_i$  within the time interval $[0, T_{slow}/2]$  for the discussed essentially two-dimensional solution.
  One can see that, during $T_{slow}/2$ time interval, the solution  passes through the following stages:
   one-dimensional solution in the form of oblique patterns with $\bm q_{c 1}=(q_{c x}, q_{c y})$ at $t_1$; an essentially two-dimensional lattice solution at $t_2$;  one-dimensional solution in the form of oblique patterns with $\bm q_{c 2}=(q_{c x}, -q_{c y})$ at $t_3$; an essentially two-dimensional lattice solution at half its period at $t_4$; and   one-dimensional solution in the form of oblique patterns with $\bm q_{c 1}$ at $t_5$, corresponding to the pattern of the respective standing wave after half a period relative to time instant $t_1$. During the next time interval, equal to $T_{slow}/2$, the solution returns to the state of an effectively one-dimensional solution in the form of oblique patterns with $\bm q_{c 1}$) corresponding to the time instant $t_1$; that is, the system returns to its initial state after a time interval $\Delta t = T_{slow}$.   The corresponding vector visualization  of the director distortion field  $\delta {\bm n}$ from Fig. \ref{psizdens} is presented in Fig. \ref{viz}.

 This behavior is precisely described within the framework of the corresponding two-dimensional (2D-) phenomenology formulated below, which yields a solution in terms of bifurcation theory \cite{And66,And67,HK81,Arn86}. From the perspective of bifurcation theory \cite{And66,And67,HK81,Arn86}, this highly non-trivial dynamical behavior, which we term the 'dancing' regime,  of two-dimensional patterns indicates that the long-time solution of the considered system of nonlinear equations is a stable limit cycle. This corresponds to a closed trajectory in phase space, which emerges in a nonlinear system near a Hopf bifurcation (i.e., a Poincar'{e}–Andronov–Hopf bifurcation).
   Let emphasize,  that in general we obtain the similar character of solutions  for the case of different Frank elastic constants.

\begin{figure}
\hskip-0.3true cm	\includegraphics[width=0.9\columnwidth]{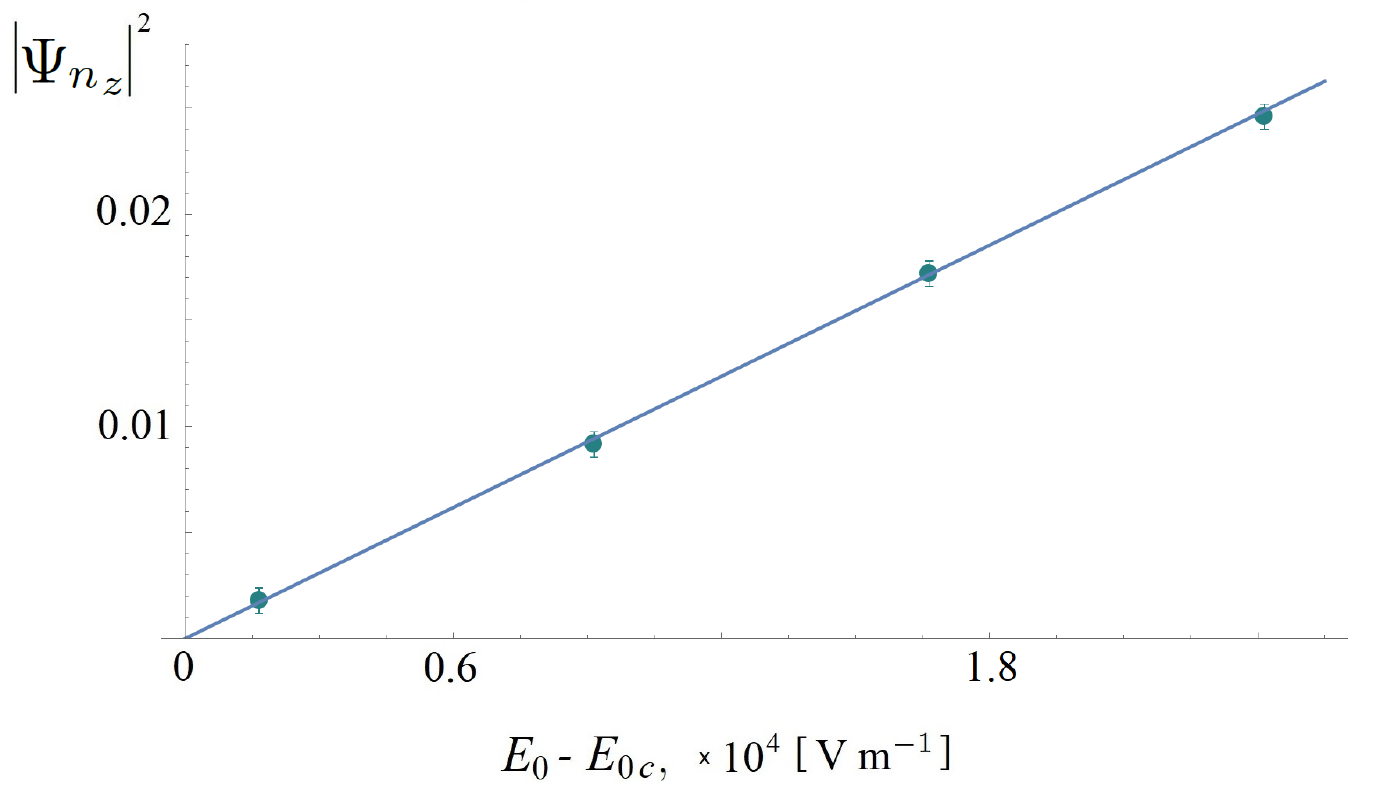}
\caption{   Dependencies  of the square of the first  harmonic   amplitude $|\Psi_{n_z}|^2$ of the  solution  for $n_z $ over $(E_0-E_{0 c})$  for the set of material parameters from the Figure \ref{psizdens}. Straight line is the linear function $\propto (E_0-E_{0c})$, fitted to the numerical blue points.  }
\label{crit}
\end{figure}

The critical character of the dependence   of the first harmonic amplitude of the solutions on the deviation of the external electric field from the threshold value $E_{0 c}$ was confirmed.
This fact is in agreement with the 2D-phenomenology and corresponds to a soft bifurcation, see Eq. (\ref{run2d3p}).
The corresponding dependence of the square   of the first harmonic amplitude of $n_z $  on  $(E_0-E_{0 c})$ is presented in Fig. \ref{crit} for the single-Frank-constant approximation for the same set of material parameters as used in Fig. \ref{psizdens}.

Note also that, with a further increase in the external field amplitude, the mode with a complex characteristic exponent (instability increment - or growth rate) — that is, the instability corresponding to oscillatory behavior — ceases to be dominant; consequently, the truly two-dimensional solution described above loses stability and becomes unstable. Thus, a hard bifurcation occurs, driving the system into a state of static straight stripes of the nematic director distortions,
 and the solution with a real characteristic exponent and  ${\bf q}_c = (q_{cx}=0,q_{cy})$ becomes the  stable one, corresponding to one-dimensional patterns with layers parallel to the $x$-axis.
However, since the transition between these competing solutions is a first-order transition, it is interesting to study hysteresis, as it is obvious that there is a possibility of long-term metastability of the solution. Hence, we find that in this case the solution in the form of a one-dimensional standing wave is a long-lived metastable solution.

In the next step, for analytical insight and interpretation of the numerical solution, it is useful to develop a complete two-dimensional phenomenological investigation. Indeed, due to the slow character of the critical mode dynamics near the instability point, it can be effectively separated from other degrees of freedom. Thus, for a weakly nonlinear regime, a Ginzburg–Landau-like 2D-phenomenological equations can be formulated, within which the emerging field structures can be described by a few parameters, see \cite{PMKL26}.
In the considered case of the slow dynamics of nematics, the solutions for the critical distortion mode related to the principal harmonic include Fourier harmonics with two equivalent tilted critical wave vectors, ${\hbox{\bf q}}_1$ and ${\hbox{\bf q}}_2$, having equal  $x$-components and opposite $y$-components.
For example the general form of the contribution to $n_z$ related to the principal harmonic can be expressed as
\begin{eqnarray}
\mathrm{Re}\, \Big\{
[\widetilde{\psi}_{1+}(t) +\widetilde{\psi}_{1-}(t)]\exp(i \bm q_1 \bm r)
\nonumber \\
+\, [\widetilde{\psi}_{2+}(t) + \widetilde{\psi}_{2-}(t)]\exp(i \bm q_2 \bm r)\Big\},
\label{runs0}
\end{eqnarray}
where $t$  is expressed  in units of $T$, $\bm q_1=\bm q_{c 1}= (q_{c x}, q_{c y})$ and $\bm q_2 =\bm q_{c 2}=(q_{c x}, - q_{c y})$ are  wave vectors of the principal harmonic; and the complex functions $\widetilde{\psi}_{j+}$ and $\widetilde{\psi}_{j-}$ have positive and negative frequencies, respectively.

The  generalized two-dimensional Ginzburg-Landau-like phenomenological theory  is developed in what follows.
The phenomenological equations must be invariant with respect to the interchange of the wave vectors  $ _1$ and $ _2$.
Thus, taking into account all the selection rules,
one can write  the following equations for the functions $\widetilde{\psi}_{j \pm}$  \cite{PMKL26}
\begin{widetext}
\begin{eqnarray}
\partial_t \widetilde{\psi}_{1+}=\lambda \widetilde{\psi}_{1+} - b_1 |\widetilde{\psi}_{1+}|^2 \widetilde{\psi}_{1+}
- b_2 |\widetilde{\psi}_{1-}|^2 \widetilde{\psi}_{1+} - b_3 |\psi_{12 +}|^2 \widetilde{\psi}_{1+}
- b_4 |\widetilde{\psi}_{2 -}|^2 \widetilde{\psi}_{1+} - b_5\,  \widetilde{\psi}_{2 +} \widetilde{\psi}^\star_{2-} \widetilde{\psi}_{1-}  \, ,
\label{02d1}  \\
\partial_t \widetilde{\psi}_{1-}=\lambda^\star \widetilde{\psi}_{1-} -b_1^\star |\widetilde{\psi}_{1-}|^2 \widetilde{\psi}_{1-}
-b_2^\star |\widetilde{\psi}_{1+}|^2 \widetilde{\psi}_{1-} -   b_3^\star |\widetilde{\psi}_{2 -}|^2 \widetilde{\psi}_{1-}
- b_4^\star |\widetilde{\psi}_{2 +}|^2 \widetilde{\psi}_{1-}  - b_5^\star\,  \widetilde{\psi}_{2 -} \widetilde{\psi}^\star_{2+} \widetilde{\psi}_{1+}  \,,
\label{02d2}\\
\partial_t \widetilde{\psi}_{2+} = \lambda \widetilde{\psi}_{2 +} - b_1 |\widetilde{\psi}_{2 +}|^2 \widetilde{\psi}_{2 +}
- b_2 |\widetilde{\psi}_{2 -}|^2 \widetilde{\psi}_{2 +} - b_3 |\widetilde{\psi}_{1+}|^2 \widetilde{\psi}_{2 +}
- b_4 |\widetilde{\psi}_{1-}|^2 \widetilde{\psi}_{2 +} - b_5\,  \widetilde{\psi}_{1+} \widetilde{\psi}^\star_{1-} \widetilde{\psi}_{2-} \,,
\label{02d3} \\
\partial_t \widetilde{\psi}_{2-} = \lambda^\star \widetilde{\psi}_{2 -} - b_1^\star |\widetilde{\psi}_{2 -}|^2 \widetilde{\psi}_{2 -}
- b_2^\star |\widetilde{\psi}_{2 +}|^2 \widetilde{\psi}_{2 -} - b_3^\star |\widetilde{\psi}_{1-}|^2 \widetilde{\psi}_{2 -}
- b_4^\star |\widetilde{\psi}_{1+}|^2 \widetilde{\psi}_{2 -}\,  - b_5^\star\,  \widetilde{\psi}_{1-} \widetilde{\psi}^\star_{1+} \widetilde{\psi}_{2 +}.
\label{02d4}
\end{eqnarray}
\end{widetext}

Because a stationary state corresponds to a time-oscillating behavior of the principal harmonic, one can find from
Eqs. (\ref{02d1})--(\ref{02d4}) that
the equations for the absolute values of  $\widetilde{\psi}_{j \pm}$  are separated from the equations for the phases.
In this case  expression (\ref{runs0})  can be conveniently written as
\begin{eqnarray}
\hskip-0.0true cm\mathrm{Re} \big\{
[\psi_{1+}\exp(i\varphi_{1 +}(t)) + \psi_{1-}\exp(-i\varphi_{1 -}(t))]\exp(i\bm q_1 \bm r)
\nonumber \\
\hskip-0.7true cm + [\psi_{2+}\exp(i\varphi_{2 +}(t)) + \psi_{2-}\exp(-i\varphi_{2 -}(t))]\exp(i \bm q_2 \bm r)\big\}\, , \quad
\label{runs1}
\end{eqnarray}
where  the  principal harmonics amplitudes  $\psi_{j +}$ and $\psi_{j -}$ correspond to the  positive and negative frequencies, respectively, $j=1,2$ and $\psi_{j \pm} >0$;  and $\varphi_{1 \pm}(t)=\lambda'' t + \phi_{1 \pm}$ and $\varphi_{2\pm}(t)=\lambda'' t + \phi_{2 \pm}$,   $\lambda'' =  |Im(\lambda)|$,
$\phi_{j \pm}$ are the corresponding phases.
Note that \cite{PMKL26} essentially addressed the phenomenology only for a phase shift equal to zero.

Then, using Eq. (\ref{runs1}), one can rewrite   Eqs. (\ref{02d1})--(\ref{02d4}) for the amplitudes $\psi_{j \pm}$ in the form
\begin{eqnarray}
\hskip-0.3true cm \partial_t \psi_{1\pm}= \mathcal{L}_{1\pm}'  \psi_{1 \pm}
 -  b_5' \, \psi_{2+} \psi_{2-}   \psi_{1\mp} \exp(\pm i (\phi_+ +  \phi_-))  \, , \ \quad
\label{2d1}  \\
  \hskip-0.0true cm \partial_t \psi_{2\pm} = \mathcal{L}_{2\pm}'  \psi_{2\pm} - b_5'\, \psi_{1+} \psi_{1-}  \psi_{2\mp}  \exp(\mp i (\phi_+ +  \phi_-))  \, , \ \quad
\label{2d4}
\end{eqnarray}
where the following functions over $\psi_{j \pm}$, $\psi_{j \pm} >0$, are introduced for description simplification
\begin{eqnarray}
\hskip-0.3true cm \mathcal{L}_{1\pm}'=\lambda'-b_1'\psi_{1\pm}^2
- b_2' \psi_{1 \mp} ^2  - b_3' \psi_{2\pm} ^2-b_4' \psi_{2\mp}^2
 , \quad
\label{simp5}\\
\hskip-0.3true cm \mathcal{L}_{2\pm}'=\lambda'-b_1' \psi_{2\pm}^2
-b_2' \psi_{2\mp} ^2-b_3'  \psi_{1\pm}^2-b_4' \psi_{1\mp}^2
\,  ,  \quad
\label{simp8}
\end{eqnarray}
where $\lambda', b_l'$ are the real parts of the quantities,  which are assumed to be positive for  indices   $l=1,...,4$. The sign of the coefficient $b_5'$, which is the real part of the $b_5$, is not constrained and remains to be determined.

We introduce the phase shifts of the time-dependent prefactors of the two
constituent principal harmonics  $ _1$ and $ _2$  as  $\phi_+\, =\,\phi_{2+}-\phi_{1+}$, $\phi_-\, =\,\phi_{2-}-\phi_{1-}$ in Eqs. (\ref{2d1})--(\ref{2d4}).
The stationary state corresponds to a time-oscillating behavior of the principal harmonics.

Because Eqs. (\ref{2d1})--(\ref{2d4})  must be invariant with respect to the interchange of the wave vectors $ _1$ and $ _2$, the possible values for the  phase shifts $\phi_{\pm}$ of the principal garmonics temporal prefactors    are $\phi=0$ and $\phi_+= \phi_- = \pm\pi/2$. The case where $\phi_+= -\,\phi_- = \pm\pi/2$  reduces to the case with $\phi=0$.

Let us consider the situation  $\phi_+= \,\phi_- = \pm\pi/2$.
One can find from
Eqs. (\ref{2d1})--(\ref{2d4}) the following conditions of realizing of the stationary state in this case
\begin{eqnarray}
\mathcal{L}_{1\pm}' \psi_{1 \pm} +  b_5' \,\psi_{2 +} \psi_{2 -} \, \psi_{1\mp}  =0  , \
\label{2d1s3}  \\
\mathcal{L}_{2\pm}' \psi_{2 \pm} + \, b_5' \, \psi_{1 +} \psi_{1 -} \, \psi_{2\mp}  =0  . \
\label{2d4s3}
\end{eqnarray}

Evidently, the system of equations (\ref{2d1s3})--(\ref{2d4s3}) possesses the high-symmetry solutions identical to those obtained in \cite{PMKL26}, corresponding to the effective one-dimensional traveling wave or one-dimensional standing wave, respectively, as well as analogous effective one-dimensional (1D) solutions obtained by an obvious interchange of indices:
\begin{eqnarray}
\psi_{1+}=\sqrt{\lambda'/b_1'}, \quad
\psi_{1-}= \psi_{2+}=  \psi_{2-}=0 \, ,
\label{run2d1}
\end{eqnarray}
corresponding to the  1D-traveling wave, and
\begin{eqnarray}
\psi_{1+}= \psi_{1-} = \sqrt{\lambda'/{(b_1'+b_2')}} \,  , \,\quad
 \,  \psi_{2+}=  \psi_{2-}=0 \, ,
\label{run2d2}
\end{eqnarray}
corresponding to the 1D-standing wave.
In addition, the following stationary solutions of  Eqs. (\ref{2d1s3})--(\ref{2d4s3}) exist:
\begin{eqnarray}
\psi_{1+}= \psi_{2+}=\sqrt{\lambda'/{( b_3'+ b_1')}} \, , \quad
\psi_{1-}=   \psi_{2-} = 0 \, , \,
\label{run2d8p}
\end{eqnarray}
 which  is two-dimensional and corresponds to two 1D-traveling waves propagating in mutually perpendicular directions.
Evidently, analogous solutions exist with an obvious interchange of indices.

Finally, the following stationary solution of Eqs. (\ref{2d1s3})--(\ref{2d4s3}) is possible
\begin{eqnarray}
\psi=\psi_{1 \pm} = \psi_{2 \pm} =  \sqrt\frac{\lambda'\,}{\big( (b_1'+b_2')  + (b_3'+ b_4' - b_5') \big)} \, , \
\label{run2d3p}
\end{eqnarray}
corresponding to the nontrivial 2D-standing   wave which is sum of two orthogonal one-dimensional standing waves with a phase shift of  $\phi_+= \phi_- = \pm\pi/2$.


In turn, stability analysis of Eqs. (\ref{2d1})--(\ref{2d4}) for the case $\phi_+= \,\phi_- = \pm\pi/2$ shows that the decrements (i.e. damping rates) $\alpha$ determining the relaxation $\propto \exp(-\alpha t)$ of small perturbations of $\psi_{1 \pm}$  and $\psi_{2 \pm}$  near the  one-dimensional traveling-wave solution (\ref{run2d1}) are the same as the corresponding expressions from \cite{PMKL26}:
\begin{eqnarray}
\alpha_1=2\lambda',
\alpha_j= \lambda'\,{(b_j'- b_1')}/{b_1'} \, , \quad \hbox{where $j=2,3,4$}
\, . \
\label{runs2da}
\end{eqnarray}
 Thus, the solution in the form of  effective 1D-traveling wave
 is stable if $b_2'>b_1'$, $b_3'>b_1'$ and $b_4'>b_1'$ for the considered case; thus, the simultaneous  realization of three such conditions is unlikely.
Similarly, stability analysis of Eqs.  (\ref{2d1})--(\ref{2d4}) for the case $\phi_+= \,\phi_- = \pm\pi/2$ shows that  decrements $\beta$ determining the relaxation $\propto \exp(-\beta t)$ of small perturbations of $\psi_{1 \pm}$  and $\psi_{2 \pm}$ near the 1D-standing wave solution, see (\ref{run2d2}),  also match the corresponding expressions from  \cite{PMKL26}, and are equal to
\begin{eqnarray}
\beta_1= 2 \lambda', \,
\beta_2= 2 \lambda'\, {(b_1'- b_2')}/{(b_1'+ b_2')} , \,
\nonumber \\
\beta_{3, 4} = \lambda'\,\big({b_3'+b_4' \mp b_5' - (b_1'+ b_2')\big)}/{(b_1'+ b_2')} \, ,\
 \label{runs2das}
\end{eqnarray}
 This means that the   1D-standing wave is stable if $b_1'>b_2'$ and $(b_3'+b_4'  - b_5')>(b_1'+ b_2')$. Thus, two conditions must be fulfilled simultaneously for stability rather than the three required for the  one-dimensional traveling wave; these two are more likely to be met.

Next, stability analysis of Eqs. (\ref{2d1})--(\ref{2d4}) for the case $\phi_+= \,\phi_- = \pm\pi/2$ shows that the  decrements $\varkappa$ determining the relaxation $\propto \exp(-\varkappa t)$ of small perturbations of $\psi_{1 \pm}$  and $\psi_{2 \pm}$ near the solution  (\ref{run2d8p}) are
\begin{eqnarray}
\varkappa_1= 2 \lambda'\,\,, \quad \varkappa_2= 2 \lambda'\,  (b_1' - b_3')/{( b_3'+ b_1')}\, , \quad
\nonumber\\
\varkappa_{3,4}= \, \lambda' \, \Big(\frac{ b_2'+ b_4'  \mp b_5' - (b_1'+ b_3')}{( b_3'+ b_1')}\Big)\, . \quad
 \label{al32p}
\end{eqnarray}
  Thus, the solution (\ref{run2d8p}) corresponding to the sum of two 1D-traveling waves propagating in mutually perpendicular directions,  is stable if $b_1'>b_3'$ and $b_2'+ b_4'  - b_5' >(b_1'+ b_3')$.

We recall that the  solutions  (\ref{run2d3p}) for the amplitudes $\psi_{1 \pm}$  and $\psi_{2 \pm}$
correspond to the 2D-standing   wave which is the sum of two  orthogonal 1D-standing waves.
In turn,  the stability analysis of equations   (\ref{2d1})--(\ref{2d4}) for the case $\phi_+= \,\phi_- = \pm\pi/2$ shows, that  decrements $\widetilde{\varkappa}$ determining the relaxation $\propto \exp(-\widetilde{\varkappa} t)$ of small perturbations of $\psi_{1 \pm}$  and $\psi_{2 \pm}$ near the solution for 2D-standing wave (\ref{run2d3p}) are
\begin{eqnarray}
\widetilde{\varkappa}_1 = \,2\,\lambda', \quad
 \label{kap1p}\\
 \widetilde{\varkappa}_2 = 2\,\lambda'\, \Big(\frac{ b_1' + b_2' + b_5' -(b_3' + b_4' )}{( b_1' + b_2' + b_3' + b_4'  - b_5' )} \Big)\,, \quad
 \label{kap2p}\\
\widetilde{\varkappa}_3 = 2\,\lambda'\, \Big(\frac{ b_1' + b_3' + b_5' -(b_2' + b_4' )}{( b_1' + b_2' + b_3' + b_4'  - b_5' )} \Big) \, , \quad
 \label{kap3p}\\
\widetilde{\varkappa}_4 =  2\,\lambda'\, \Big(\frac{ b_1' +  b_4'  + b_5'-  (b_2' + b_3')}{( b_1' + b_2' + b_3' + b_4' - b_5' )} \Big)   \, . \
\label{kap4p}
\end{eqnarray}
One can see that solutions  (\ref{run2d3p}) are stable if
\begin{eqnarray}
 b_1' + b_2' + b_5' -(b_3' + b_4' )\, >0 \,, \quad
 \label{nkap2p}\\
 b_1' + b_3' + b_5' - (b_2' + b_4' ) \, >0  \, , \quad
 \label{nkap3p}\\
 b_1' +  b_4'  + b_5' -  (b_2' + b_3')\, >0   \, . \
\label{nkap4p}
\end{eqnarray}

A comparative analysis of the stability conditions for all solutions shows that conditions (\ref{nkap2p})--(\ref{nkap4p}) are the most likely to be satisfied for positive $b_5'$.
Thus, we find that the two-dimensional solution (\ref{run2d3p})  with phase shift  $\phi_+= \phi_- = \pm\pi/2$ is more likely in this case.

We emphasize that for a zero phase shift ($\phi=0$), all equations and resulting expressions are obtained from their counterparts with $\phi_+= \,\phi_- = \pm\pi/2$ by reversing the sign of the parameter  $b_5'$.
We also emphasize that all damping rates obtained above are of the order of $\lambda'$. As $\lambda'\propto E_0-E_{0 c}$, this means  that all instability decrements, increments  are proportional to $E_0-E_{0 c}$.

Note that we must also take into account phase distortions.
Let us consider the  case of the stationary solutions  (\ref{run2d3p}), when
$|\widetilde{\psi}_{1+}|= |\widetilde{\psi}_{1-}| = |\widetilde{\psi}_{2 +}|= |\widetilde{\psi}_{2 -}| = \psi $, $\phi_+^{(0)}= \phi_-^{(0)} = \pi/2$.
Let us introduce  the phase distortions  $\phi_+= \pi/2 + \delta\phi_+\,, \, \phi_- = \pi/2 + \delta\phi_-$, where
\begin{eqnarray}
\phi_{1+}(t) = \phi_{1+} + \delta\phi_{1+}\,\, , \quad \phi_{1-}(t) = \phi_{1-} + \delta\phi_{1-}\,\, , \
\label{phampl1} \\
\phi_{2+}(t) = \phi_{2+} + \delta\phi_{2+}\,\, , \quad \phi_{2-}(t) = \phi_{2-} + \delta\phi_{2-}\,\, , \
\label{phampl2}
\end{eqnarray}
$\phi_+^{(0)}=\phi_{2+}-\phi_{1+}$,$\phi_-^{(0)}=\phi_{2-}-\phi_{1-}$, $\delta\phi_+ = \delta\phi_{2+}- \delta\phi_{1+}$ and $\delta\phi_- = \delta\phi_{2-}- \delta\phi_{1-}$.
Then one can obtain  the following equations for $\delta\phi_{+}$ and $\delta\phi_{-}$ from
Eqs. (\ref{02d1})--(\ref{02d4}), using expressions (\ref{runs1}), (\ref{phampl1})--(\ref{phampl2})
\begin{eqnarray}
\partial_t \delta\phi_{+} =  -\,2\, |\psi|^2 \,  b_5'\,\sin(\delta\phi_+ + \delta\phi_-)\, ,
\label{2d1s2ph1}  \\
\partial_t \delta\phi_{-} = \,2\, |\psi|^2 \,  b_5'\,\sin(\delta\phi_+ + \delta\phi_-) \, .
\label{2d4s2ph2}
\end{eqnarray}
And it follows from Eqs. (\ref{2d1s2ph1})--(\ref{2d4s2ph2}) that
\begin{eqnarray}
\partial_t ( \delta\phi_{+} + \delta\phi_{-})= 0 \, .
\label{2d4s2ph3}
\end{eqnarray}
Thus we can use the solution
\begin{eqnarray}
  \delta\phi_{+} + \delta\phi_{-}= 0 \, , \
\label{2d4s2ph4}
\end{eqnarray}
 which means  that, for a weakly nonlinear regime under an external alternating electric field slightly above the flexoelectric instability threshold, the dynamic equations for the phases are decoupled from the equations for the amplitudes. Consequently, the previously obtained stability conditions are correct.
It should be stressed that for the equations governing phase distortions as well, in the case of a zero phase shift ($\phi=0$), all equations and resulting expressions are obtained from the corresponding formulas for a nonzero phase shift $\phi_+= -\,\phi_- = \pm\pi/2$ by reversing the sign of the parameter $b_5'$.

Thus, in the weakly nonlinear regime, for $b_1'+b_4'+b_5'-(b_2'+b_3' )>0$ and under the conditions $b_1' + b_2' + b_5' > (b_3' + b_4' )$ and $b_1' + b_3' + b_5' >(b_2' + b_4' )$, we conclude from the stability analysis of Eqs.  (\ref{2d1})--(\ref{2d4}) for the case $\phi_+= \,\phi_- = \pm\pi/2$, that for $b_5'>0$, the two-dimensional solution (\ref{run2d3p}) with the phase shift $\phi_+= \phi_- = \pm\pi/2$ is stable. In  opposite case $b_5'<0$ the two-dimensional solution with $\phi_+= \phi_- = 0$ is stable.

 Given that numerical calculations show the nontrivial two-dimensional solution with principal harmonics amplitudes (\ref{run2d3p}) and the phase shift of the principal harmonics $\phi_+= \phi_- = \pm\pi/2$ to be stable for the case of a complex characteristic exponent, this implies that $b_5'>0$ is more probable in this case.
Stability analysis of the phenomenological equations  (\ref{2d1})--(\ref{2d4}) for the case $\phi_+= \,\phi_- = \pm\pi/2$ shows the same result.
In this case, the   stationary solution  of phenomenological Eqs. (\ref{2d1s3})--(\ref{2d4s3}) for the amplitudes $\psi_{j \pm}$ given by Eq. (\ref{run2d3p})  is stable.
The regime corresponds to the nontrivial two-dimensional   wave in the form of 'dancing' two-dimensional  patterns, described above, when
 expression  (\ref{runs1}) corresponds to two orthogonal one-dimensional standing waves with a phase shift of  $\phi_+= \phi_- = \pm\pi/2$.

 To describe the result two-dimensional solution  and relate it with the developed phenomenology, it is advantageous to isolate the time-dependent prefactors of the two constituent principal harmonics modulating the principal harmonics' spatial profiles in Eq. (\ref{runs1}), and to define the absolute values of the prefactors as
 \begin{eqnarray}
  \Psi_1(t) =  \vert\psi_{1+}\exp(i\varphi_{1 +}(t)) + \psi_{1-}\exp(-i\varphi_{1 -}(t)) \vert\, ,
   \nonumber \\
    \Psi_2(t) =  \vert\psi_{2+}\exp(i\varphi_{2 +}(t)) + \psi_{2-}\exp(-i\varphi_{2 -}(t)) \vert \, . \ \
  \label{psi}
\end{eqnarray}
 The corresponding dependencies of  $\Psi_{n_z 1}$ and $\Psi_{n_z 2}$ for the $n_z$,
   obtained from the numerical solution of the  nonlinear dynamic equations  (\ref{trun4})--(\ref{pres}), on time during the time interval from Fig.\ref{psizdens} are shown in Fig.\ref{psiz}.
 These dependencies  illustrate the two-dimensional solution behavior with a phase shift of  $\phi_+= \phi_- = \pm\pi/2$ between its constituent orthogonal one-dimensional standing waves   described above.  For each given electric field  deviation $E_0-E_{0 c}$ we have $\Psi_{n_z}^2 = \Psi_{n_z 1}^2(t) + \Psi_{n_z 2}^2(t)$, where $\Psi_{n_z}$ is equal to multiplied by 2 amplitude $\psi$ determined by  expression (\ref{run2d3p}), i.e. equal to $2\psi$. This behaviour is in agreement with properties of the  limit cycle.  One can term $\Psi_{n_z}^2$ as square of the first  harmonic   amplitude.

To connect phenomenological  expression (\ref{runs1}) with the numerical two-dimensional solution obtained above, we examine the evolution within the time interval $[0, T_{slow}/2]$ (Fig. \ref{psizdens})  at the same five  time instants $t_i$. At $t=t_1$  the 1D standing wave with wave vector $\bm q_1$ and time-dependent prefactor absolute value equal to $2\psi$,  dominates and time-dependent prefactor  of the 1D standing wave with wave vector $\bm q_2$ is equal to 0 and we have solution in the form of one dimensional standing wave in main approximation. At  $t=t_2$ both ortogonal 1D standing waves have equal time-dependent prefactor absolute value equal to $\sqrt{2} \psi$ and we have two dimensional solution. At  $t=t_3$   the 1D standing wave with wave vector $\bm q_1$  vanishes,  leaving 1D standing wave with wave vector $\bm q_2$ and time-dependent prefactor absolute value equal to $2\psi$  dominated.  At $t=t_4$ both standing waves have equal time-dependent prefactor absolute value equal to $\sqrt{2} \psi$, forming a 2D state.  Finally, at $t=t_5$ the 1D standing wave with wave vector $\bm q_1$ and time-dependent prefactor absolute value equal to $2\psi$ dominates and the standing wave with wave vector $\bm q_2$  vanishes. It recovers the initial 1D configuration with solution in the form of 1D standing wave 1 but phase-shifted over $\pi$ relative to time instant $t_1$. During the next time interval, equal to $T_{slow}/2$, the solution returns to the state at $t=t_1$.
 Accordingly we  conclude that the numerical results are in agreement with the developed Ginzburg-Landau-like two-dimensional phenomenology near the instability threshold.

 We emphasize that the critical behavior of $\lambda'$ near the threshold value of the external field amplitude $E_{0 c}$ follows the law  $\lambda'\propto E_0-E_{0 c}$. Therefore the amplitudes of $\psi_{i \pm}$ (see (\ref{run2d3p}) are proportional to $\sqrt{E_0-E_{0 c}}$,  thus ensuring the critical behavior of the dependence of the square of the first harmonic amplitude,  $|\Psi_{n_z}|^2$, presented in Fig. \ref{crit}. In turn   $\lambda''$  is determined  by the imaginary part of $\lambda$. But as we obtained in \cite{PK24} the values of $Im(\lambda)$, and correspondingly $\lambda''$ weakly depend on the electric field
in the vicinity of the threshold, i.e. they weakly depend on the deviation of $E_{0}$ from its threshold value $E_{0 c}$. It means, that the slow  dynamics  frequency of the solution has no critical behavior. Note that only  corrections to the frequency related to the imaginary parts of $b_i$ are proportional  to $E-E_c$, see Eqs. (\ref{02d1})--(\ref{02d4}).

Thus, the present work demonstrates that, in a weakly nonlinear regime where the material parameters and external field amplitude values correspond to the region of the oscillating regime in the linear approximation, the genuinely two-dimensional solution with principal harmonics amplitudes  (\ref{run2d3p}) and  a phase shift of  $\phi_+= \phi_- = \pm\pi/2$ between its constituent orthogonal one-dimensional standing waves is stable.   This  solution has the widest basin of attraction.
With this numerical result in hand, one can determine the necessary conditions for the phenomenological constants from the presented two-dimensional Ginzburg–Landau–like phenomenology to yield such a solution.
Solutions in the form of one-dimensional standing waves are realized only under initial conditions close to a one-dimensional standing wave.
 This fact indicates a narrow basin of attraction and metastability for such a solution.

The current study provides a striking  example of nonlinear dynamical system behavior in nematics.
  By solving the  set of exact nonlinear equations, derived in \cite{PM23,PMKL26},  we demonstrate the emergence of a 'dancing' behavior of two-dimensional patterns above the flexoelectric instability threshold.
 The analysis considers liquid crystals with negative dielectric and conductivity anisotropies.
 Just above the Hopf bifurcation, corresponding to a complex characteristic exponent, the solution represents a stable limit cycle of the nonlinear system.
 The problem of the regimes and character of solutions in the case of external field amplitudes significantly exceeding the threshold value at the Hopf bifurcation is the subject of further ongoing research.

The type of flexoelectric instability presented here may find useful applications in integrated photonics, in particular for the development of next-generation optical communication systems, see Refs. \cite{Ert23,IS16}.
   When planar waveguides are filled with liquid crystals \cite{Wh77,Wei20} possessing the described properties, and alternating field sources are placed at specific locations—integrated within optical add-drop multiplexers \cite{HB13,As06} with polarizers—such a device will transmit a selected type of signal at predicted time intervals, which is a temporal analog to spatial filtering logic based on polarization \cite{LW87}.
   Specifically, it will pass light with one of four polarization types through four corresponding time intervals.
   The discussed procedure could potentially also be used for signal encoding.

\acknowledgments

I am sincerely grateful to V.V. Lebedev and E.I. Kats  for countless fruitful discussions on the problems of the nonlinear physics and nematodynamics.


The work of was supported  by the Ministry of Science and Higher Education of the Russian Federation (State assignment No. FFWR-2024-0014).

\end{document}